\documentclass{aa}  

\usepackage{graphicx}
\usepackage{subcaption}
\usepackage{tabularx}
\usepackage{txfonts}
\usepackage{appendix}
\usepackage{float}
\usepackage{caption}
\begin{document}

   \title{The bimodality in the mass-metallicity relation in SDSS-MaNGA galaxy pairs}


   \author{Kiyoaki Christopher Omori
          \inst{1}
          \and
          Tsutomu T. Takeuchi\inst{1, 2}
          }

  \institute{$^1$Division of Particle and Astrophysical Science, Nagoya University, 
Furo-cho, Chikusa-ku, Nagoya 464--8602, Japan\\
$^2$The Research Centre for Statistical Machine
Learning, the Institute of Statistical Mathematics, 10--3 Midori-cho, Tachikawa, Tokyo 190--8562, Japan
}

   \date{Received September 15, 1996; accepted March 16, 1997}

 
  \abstract
   {}
   {Interacting galaxies show a metallicity dilution compared to isolated galaxies of similar masses in the mass-metallicity space at the global scale.
   We investigate the spatially resolved mass-metallicity relation (MZR) of galaxy pairs in the SDSS-MaNGA survey to confirm that the local relation between the stellar mass surface density, $ \Sigma_{*}$, and the metallicity is consistent with the MZR at the global scale.}
   {We investigate the relationship between the stellar mass surface density and the metallicity abundance, $12 + \log \text{(O/H),}$ for star-forming spaxels belonging to 298 galaxy pairs identified using visual and kinematic indicators in the SDSS-MaNGA survey. We also investigate if \textbf{a)} the location of a spaxel relative to the galaxy centre and \textbf{b)} the galaxy pair separation have any effect on the local MZR.}
   {We find that the correlation between mass and metallicity holds for interacting galaxies at the local level. However, we find two peaks in spaxel distribution, one peak with enriched metallicity and the other with diluted metallicity. We find that the spaxels belonging to the galaxy central regions (i.e. at lower $R/R_{\textrm{eff}}$) are concentrated close to the two peaks. We also find that the metallicity-diluted spaxels belong to galaxy pairs with closer projected separations and that spaxels with enriched metallicity belong to galaxy pairs with greater projected separations. }
   {We find two discrete peaks in the spatially resolved MZR for star-forming spaxels that belong to galaxy pairs. The peaks are likely related to the galaxy projected separation or the stage of the interaction process of a galaxy pair.}


   \maketitle
%

\section{Introduction}

    In the current $\Lambda$-dominated cold dark matter ($\Lambda$CDM) framework for structure formation in the Universe, the hierarchical growth of galaxies through merging is the commonly agreed upon pathway for galaxy evolution.
    Despite galaxy interactions and mergers being a major driver of galaxy evolution, we are yet to make many quantitative conclusions on the process, and our understanding of galaxy interactions and mergers is far from complete.
    One such example is the chemical evolution of interacting galaxy systems.
    While the chemical evolution of isolated galaxies has been extensively studied, there have been comparatively fewer studies done on the topic for interacting galaxies.
    
    We can improve our understanding of chemical evolution in interacting galaxies by examining the relationship between two fundamental properties of galaxies: stellar mass ($M_*$) and gas phase metallicity (hereinafter referred to as `metallicity').
    Stellar mass can be an indicator of the amount of gas converted into stars during a galaxy's lifetime.
    Metallicity can be a tracer of gas reprocessed into stars or accreted due to external processes and can be a reflection of the state of galaxy evolution.
    Both of these properties can change as a consequence of star formation events.
    The relationship between these two properties is called the mass-metallicity relation (MZR; \citealt{1979A&A....80..155L}). 
    This relationship indicates that the metallicity, in particular the oxygen abundance, of galaxies increases with increasing stellar mass \citep{2004ApJ...613..898T, refId0}.
    Lower-mass galaxies are more greatly affected by blowouts due to galactic winds or outflows from galactic processes; this results in metal content leaving the galaxy, which in turn dilutes the gas phase metallicity.
    On the contrary, higher-mass galaxies are more chemically enriched. This could be due to higher-mass galaxies being less affected by the above processes and having the ability to retain their metal content, or it could be a consequence of `chemical downsizing' \citep{2015ARA&A..53...51S}.
    At larger stellar masses, the relation bends and flattens off towards an asymptotic value. This behaviour indicates some sort of saturation value for gas phase metallicity, possibly due to the galactic outflow, which regulates metallicity \citep{2004ApJ...613..898T}.
    
    This relation has been shown to also hold true at a local, or spatially resolved, scale.
    Studies of HII regions for a small number of galaxies have shown that a relationship exists between stellar mass density and metallicities, with denser regions exhibiting a greater metallicity \citep{1984MNRAS.211..507E, 1992MNRAS.259..121V}.
    Later, \citet{2012ApJ...745...66M} found a local correlation between stellar mass density and metallicity using long-slit spectra of galaxies in the Galactic All Sky Survey (GASS; \citealt{2009ApJS..181..398M}).
    The advancement in integral-field-spectroscopy (IFS) techniques has allowed for detailed analyses on the spatially resolved properties of a larger sample of galaxies.
    Studies using spatially resolved spectra have confirmed that the relation between the stellar mass density and metallicity is a locally scaled version of the global MZR \citep{2012ApJ...756L..31R, 2013A&A...554A..58S, 2016MNRAS.463.2513B}.

    Recent studies have shown that the metallicity of galaxy mergers falls below the MZR, or in other words, interacting galaxies have lower nuclear metallicities than those of isolated galaxies of similar masses  \citep{2008AJ....135.1877E, 2012MNRAS.426..549S, 2017MNRAS.467.3898C}.
    In particular, close galaxy pairs show an offset from the MZR.
    A likely explanation for this is gas inflow to galaxy core regions during a merger event \citep{Rupke_2010, Montouri2010, 2011MNRAS.417..580P}.
   The dilution in metallicity due to this inflow event occurs at a galaxy-wide scale and is not just a local phenomenon occurring at the galaxy centre \citep{2018MNRAS.480.2544R}.
    Accreted lower-metallicity gas from a galaxy merger will flow into a higher-metallicity central region, resulting in a lower gas phase metallicity.
    
    In this paper we investigate the spatially resolved MZR of galaxy pairs in the IFS survey Mapping Nearby Galaxies at Apache Point Observatory (MaNGA; \citealt{2015ApJ...798....7B}) to study the effects of interaction on a local scale. 
    We compare the loci of the spaxels of our sample to the MZR curve derived from all star-forming spaxels in the MaNGA survey to confirm if metallicity dilutions occur for galaxy pairs.
    We also investigate if the distribution of spaxels of the paired sample is affected by other parameters.
    
    For this paper, we adopt a $\Lambda$CDM model with the following cosmological parameters: $H_0 = 70\textrm{ km } \textrm{s }^{-1} \textrm{ Mpc}^{-1}$, $\Omega_{\Lambda}=0.7$, $\Omega_{M}=0.3$.
    
    The paper is structured as follows. Section \ref{section:Data} describes our sample and methods to obtain our properties. We highlight our results in Sect. \ref{section:Results}. We discuss our results in Sect. \ref{section:Discussion}, and our conclusions are presented in Sect. \ref{section:Conclusion}.

\section{Data and analysis}
\label{section:Data}

\subsection{Sample}
For this work we used data from the IFS survey MaNGA, one of the three core projects of Sloan Digital Sky Survey IV (SDSS-IV; \citealt{2017AJ....154...28B}). 
It uses the 2.5 meter telescope at the Apache Point Observatory \citep{2006AJ....131.2332G} and aimed to map and acquire spatially resolved spectroscopic observations of $\sim$10,000 local galaxies in a redshift range of 0.01 $<$ \textit{z} $<$ 0.15, with an average redshift of 0.037 \citep{Law_2016}, by 2020.
MaNGA spectra cover a wavelength range of 3,600{\AA}\mbox{--}10,000{\AA} at a resolution of R $\sim$ 2,000. 

The MaNGA target selection is optimised in such a way that galaxies are selected based only on their SDSS $i$-band absolute magnitude and redshift, and the sample is unbiased based on their sizes or environments. 
The methodology and extensive efforts taken for this optimisation are highlighted in \citet{2017AJ....154...86W}. 
We used data from SDSS Data Release 16 (DR16), which includes the spatially resolved maps of 4675 unique MaNGA targets.

\subsection{Selection}
We selected our sample of galaxy pairs using a method that combines visual identification of MaNGA cutout images, visual inspection of 2D kinematic maps, and relative velocity differences. 

 First, we visually investigated the 2D stellar kinematic maps of MaNGA galaxies. The stellar kinematic maps were obtained from the output of the data analysis pipeline (DAP) in MaNGA \citep{Westfall_2019}. 
In the DAP, the Voronoi binning method of \citet{2003MNRAS.342..345C} is used to bin, stack, and average the spectra of adjacent spaxels such that the target minimum signal-to-noise ratio (S/N) to obtain accurate stellar kinematics is met, which in this case is 10. 
The stellar continuum of each binned spectrum was fitted using the penalised pixel-fitting method by \citet{2017MNRAS.466..798C} and hierarchically clustered Medium-resolution Isaac Newton Telescope library of empirical spectra stellar library (MILES stellar library: \citealt{2006MNRAS.371..703S}). 
The stellar kinematic information (velocity and velocity dispersion) was obtained through this fitting process. 
From the 2D kinematic maps, we visually identified 1569 galaxies with disturbed stellar kinematics.

We next inspected whether or not these 1569 galaxies were galaxy pairs or isolated galaxies through visual confirmation of their optical images and SDSS galaxy pair data.
We investigated the MaNGA image cutouts of the galaxies and considered the galaxy a galaxy pair if it met one of the following two criteria:
a) a secondary galaxy was within the cutout and b) a secondary galaxy exists within the range of the SDSS Neighbours Table.

We confirmed that these galaxy pairs were within a physically connected range and not projections following the redshift difference range adopted in \citet{Patton_2000}.
After these steps, our final sample consisted of 298 galaxy pairs lying in the redshift range 0.013 $<$ \textit{z} $<$ 0.15.

\subsection{Obtaining physical properties}
After the sample selection, the next step was to extract the physical properties from the spaxels of our sample galaxies.
The properties of interest in this work are the stellar surface mass densities and gas phase metallicities. 
\subsubsection{Stellar surface mass densities}
The spatially resolved stellar surface mass densities were obtained by finding the ratio between the stellar mass and surface area of each spaxel.

To obtain the stellar mass, we referred to the MaNGA FIREFLY Value Added Catalogue (MaNGA FIREFLY VAC; \citealt{2017MNRAS.466.4731G}), which provides spatially resolved stellar population properties for MaNGA galaxies.
The MaNGA FIREFLY VAC summarises the results of running the full spectral-fitting code FIREFLY \citep{2017MNRAS.472.4297W} on spatially resolved MaNGA spectra that are binned using the Voronoi binning method with a S/N of 10 per pixel.
Details on the fitting process and how the stellar population properties are obtained are detailed in \citet{2017MNRAS.466.4731G}.

After the stellar masses for each spaxel were obtained, the surface mass densities were obtained.
In MaNGA data, each spaxel has a size of 0.5 arcsec. 
We used the small angle approximation to estimate the physical scale of the spaxel:
\begin{equation}
    \theta=\tan^{-1} \left(\frac{d}{D}\right) \approx \frac{206,625 \text{ [arcsec]}}{1 \text{ [radian]}} \frac{d}{D}
,\end{equation}
with $\theta$ the angular size of the spaxel in arcseconds, $D$ the angular diameter distance, and $d$ the diameter of the spaxel.

We approximated the distance using the Hubble law,
\begin{equation}
    D\approx\frac{cz}{H_0}
,\end{equation}
with the redshift information for each galaxy available through the DAP catalogue.
We obtained the physical scale, $\frac{d}{\theta}$, of each spaxel using the small angle approximation and then converted the spaxel size from arcseconds to parsecs to obtain the spaxel area.

We found the ratio between the stellar mass and spaxel area to obtain the surface mass density.
We corrected for any projection effects and inclination by multiplying the stellar mass surface density by the value $b/a$, with $b$ and $a$ representing the projected semi-major and semi-minor axes of the galaxy, respectively. This value was obtained from the DAP: 
\begin{equation}
    \Sigma_{*}=\frac{M_{*}}{pc^2}\frac{b}{a}
.\end{equation}
\subsubsection{Gas phase metallicities}
For this work we adopted the O3N2 metallicity calibrator from \citet{2013A&A...559A.114M} since it is also used in \citet{2016MNRAS.463.2513B}, a work that handles the spatially resolved MZR for MaNGA galaxies; 
\begin{equation}
    12 + \log \text{(O/H)} = 8.533[\pm0.012]-0.214[\pm0.012]\times \mbox{O3N2}
.\end{equation}
The O3N2 calibrator was determined by taking the logarithmic differences between the line ratios $\log(\mathrm{OIII}/\text{H}\beta)$ and $\log([\mathrm{NII}]/\mathrm{H}\alpha)$:
\begin{equation}
    \mathrm{O3N2}=\log \left({\frac{[\text{O III}]\lambda5007}{\mathrm{H}\beta}\times\frac{\mathrm{H}\alpha}{[\text{N II}]\lambda6585}}\right)
.\end{equation}
We note that diffused ionised gas (DIG) can significantly affect emission line fluxes and, consequently, the emission line ratios obtained from these fluxes \citep{1985ApJ...294..256R, 1987ApJ...323..118R, 1994ApJ...431..156W, 1996AJ....112.1429H, 1998ApJ...506..135G, 2003ApJ...586..902H, 2006ApJ...652..401M, 2006ApJ...644L..29V}.
When using the O3N2 calibrator to calculate metallicity, DIG can cause a scatter in metallicity measurements, with the metallicity appearing higher in some galaxies and lower in others \citep{2017MNRAS.466.3217Z}.  
However, offsets introduced by DIG are cancelled out when the sample size is sufficiently large \citep{2017MNRAS.466.3217Z}, an observation that we adopt in this paper.

\begin{figure}[h]
    \centering
    \begin{tabularx}{\textwidth}{X}
    \includegraphics[width=\columnwidth,keepaspectratio]{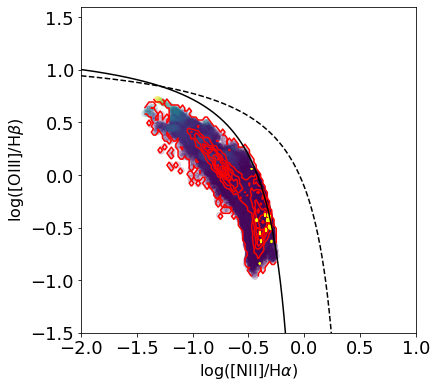}
    \end{tabularx}
    \caption{Distribution of the spaxels of our sample of galaxy pairs shown in an [OIII]/H$\beta$ and [NII]/H$\alpha$ diagnostic diagram. The solid and dotted black lines are the star-forming and composite classification lines, respectively, defined in \citet{2006MNRAS.372..961K}. The red and yellow contours represent the distribution of spaxels with H$\alpha$ equivalent width greater than or less than $6\AA$, respectively.}
    \label{fig:samplebpt}
\end{figure}

Once the surface mass densities and gas phase metallicities of all spaxels were obtained, we selected only the star-forming spaxels of each galaxy as gas phase metallicity calibrators are only accurate for star-forming spaxels.
The selection was done by comparing the [OIII]/H$\beta$ and [NII]/H$\alpha$ line ratios in a Baldwin-Phillips-Terlevich diagram (BPT diagram; \citealt{1981PASP...93....5B}).
Next, the S/N was obtained for the star-forming spaxels by multiplying each spaxel's flux by the square root of its inverse variance. {Details on the process used to obtain the inverse variance are given in \citet{Westfall_2019}.}
Of the star-forming spaxels, we left out spaxels with a S/N $<3$ in both N[II]6585 and H$\alpha$, as well as any other spaxels that lacked coverage, had unreliable measurements, or were otherwise considered in the MaNGA catalogue to be unusable for science.
We show the spaxels used in our study in Fig. \ref{fig:samplebpt}. We note that the majority of our spaxels ($~99\%$) have an H$\alpha$ equivalent width $> 6\AA$, an additional threshold for classifying star-forming spaxels \citep{2010MNRAS.403.1036C}.

\section{Results}
\label{section:Results}

\begin{figure*}[h]
    \centering
    \begin{tabularx}{\textwidth}{X}
    \includegraphics[width=\textwidth,keepaspectratio]{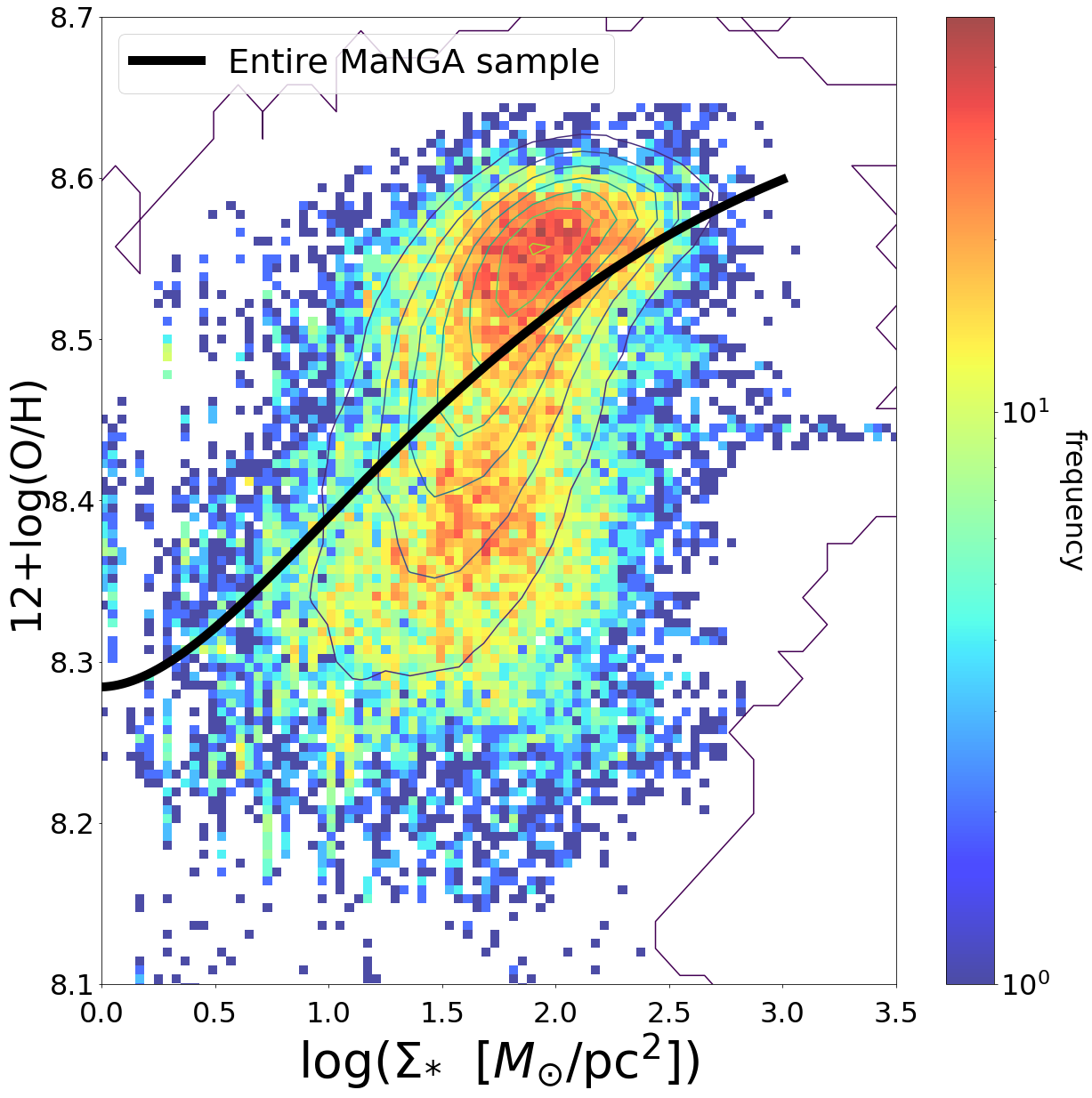}
    \end{tabularx}
    \caption{Distribution of oxygen abundance as a function of stellar mass surface density for all star-forming spaxels in the 298 galaxy pairs in the MaNGA survey. The colour bar indicates the number of spaxels per bin in the $\Sigma_{*}-\mathrm{Z}$ space. The black curve is the best-fit line found from the entire MaNGA sample. The contours indicate the distribution of all MaNGA spaxels.}
    \label{fig:samplemzr}
\end{figure*}

Figure \ref{fig:samplemzr} plots the oxygen abundance of the star-forming spaxels of our galaxy pair samples as a function of stellar mass surface density within 1.5 $R_{\textrm{eff}}$.
The same plot for all star-forming spaxels in the MaNGA survey, which we used to plot the red line in Fig. \ref{fig:samplemzr}, is available in Appendix \ref{appendix: A} as reference.

We find a bimodality in the metallicity distribution at higher stellar masses, which indicates that two populations of spaxels exist.
Such a discrete bimodality is not present in the MaNGA population as a whole, however, as seen in Appendix \ref{appendix: A}.

One of the peaks agrees with the MZR, meaning that higher stellar masses exhibit higher metallicity.
This indicates that for some galaxies pairs, the MZR at the local level is in agreement with that of the global level.

There is a secondary peak located below the fit curve for all MaNGA galaxies, albeit in agreement with the contour lines, indicating that there are spaxels with lower metallicities than those with similar stellar masses. These diluted spaxels are in relative accordance with the conclusions of studies that investigated  the MZR for galaxy pairs at a global level (e.g. \citealt{Rupke_2010, 2020MNRAS.494.3469B}), which find that galaxy pairs show a metallicity decrement compared to isolated galaxies of similar stellar mass. 
\section{Discussion}
\label{section:Discussion}

In this section we discuss the properties and possible origins of the bimodality.

There have been a number of works that discuss the MZR in interacting galaxies \citep{10.1111/j.1745-3933.2008.00466.x, Rupke_2010} and a further few that include the star formation rate and discuss the fundamental metallicity relation (FMR) in interacting galaxies \citep{Mannucci_2010, 2014MNRAS.444.3986R, 10.1093/mnras/stv1232, 2020MNRAS.494.3469B}.
In particular, \citet{2020MNRAS.499.4370M} conducted a thorough study on the spatially resolved MZR in star-forming regions of perturbed galaxies in the  Calar Alto Legacy Integral Field Area (CALIFA; \citealt{2015A&A...576A.135G}) survey.
They find that tidally perturbed galaxies show lower oxygen abundances compared to similar-mass non-perturbed galaxies.
We compared our findings with these results.
Our results in Fig. \ref{fig:samplemzr} show that while some spaxels in galaxy pairs show a dilution in gas phase metallicity, which is in agreement with the above-mentioned works that investigate the MZR for galaxy pairs, there are also spaxels that do not show diluted metallicities.
This bimodality in distribution is not present in \citet{2020MNRAS.499.4370M}.
We looked to see if there were any properties that may show a close relation with the loci of the spaxels. In this work we focus on two properties: \textbf{a)} the effective radius of each spaxel and \textbf{b)} the separation of the galaxy pair each spaxel belongs to.

\subsection{Spaxel distribution by effective radius}
\begin{figure*}[htp]
    \centering
    \begin{tabularx}{\textwidth}{X}
    \includegraphics[width=\textwidth,keepaspectratio]{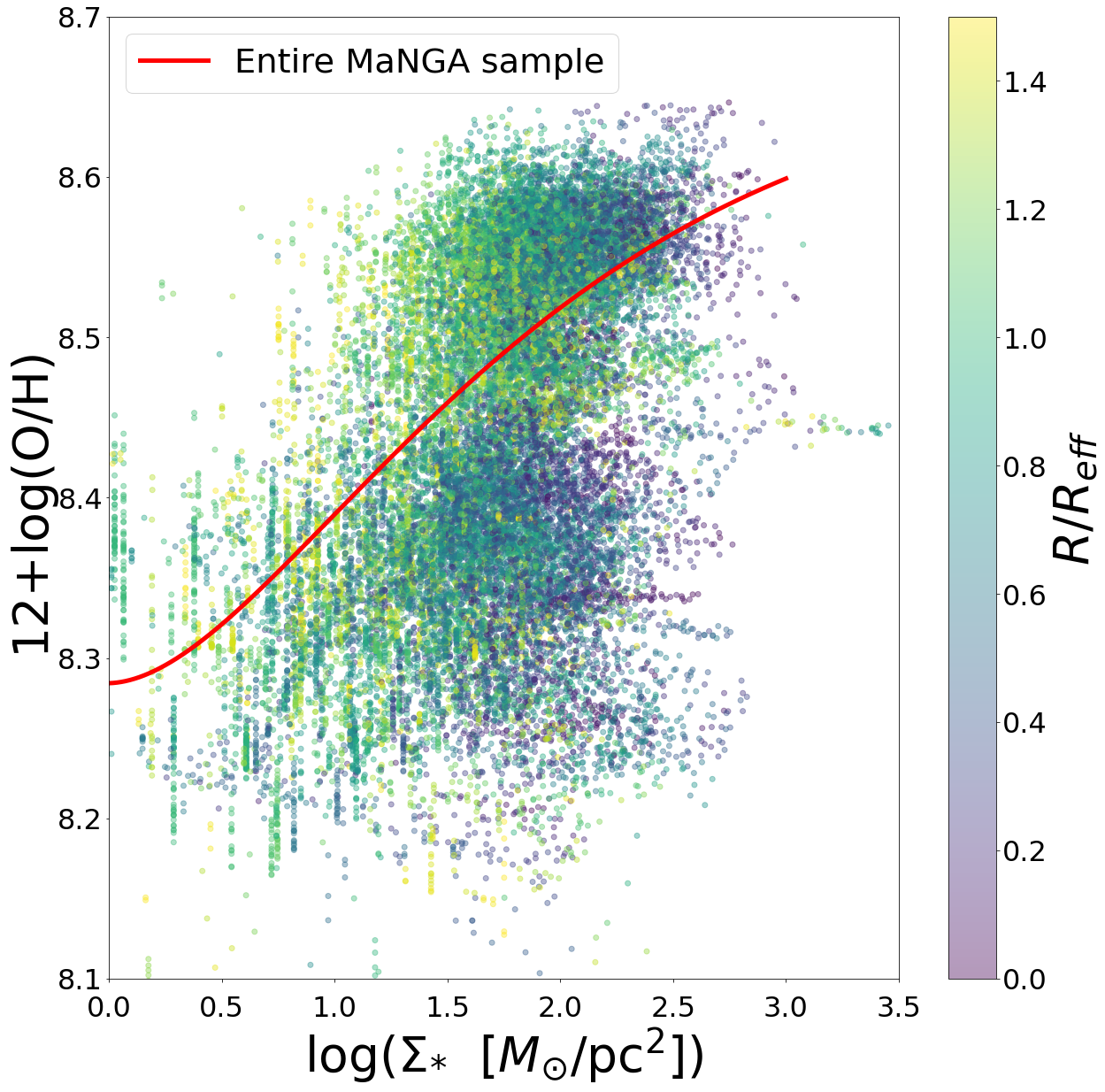}
    \end{tabularx}
    \caption{Distribution of oxygen abundance as a function of stellar mass surface density for all star-forming spaxels in the 298 galaxy pairs in the MaNGA survey, colour-coded by effective radius. All spaxels over 1 $R_{\textrm{eff}}$ are masked.}
    \label{fig:effrad}
\end{figure*}

Figure \ref{fig:effrad} shows the spaxels colour-coded by $R_{\textrm{eff}}$, extending out to 1.5 $R_{\textrm{eff}}$.
We find that the two peaks in Fig. \ref{fig:samplemzr} are similar in loci with the core regions of the galaxy.
This indicates that the central regions of galaxies contribute to the bimodality.
This bimodality indicates that there are two populations of galaxies in the sample, one population with enriched metallicity and one with a dilution of metallicity. 
The dilution in the latter population is likely a consequence of the galaxy interactions.
In a galaxy interaction, strong inflows of gas from the paired galaxy occur, and the accreted gas flows towards the circumnuclear regions of the primary galaxy, fuelling star formation \citep{2004ApJ...616..199I}.
If the paired galaxy has outer regions that are metal poor, such as in local galaxies \citep{doi:10.1146/annurev.aa.28.090190.002521}, the accreted gas will also be metal poor, resulting in lower-metallicity gas diluting the metallicity in the core region of the primary galaxy \citep{Montouri2010}.

The former population, or the higher-mass, higher-metallicity spaxels belonging to galaxy core regions, can be explained by the inside-out galaxy evolution model \citep{1999ApJ...520...59K}.
In this model, a negative metallicity gradient is observed, with the greatest metallicity values in the galaxy cores \citep{1988MNRAS.235..633V, 1992MNRAS.259..121V}.

While the effective radius indicates that there are both higher-metallicity and lower-metallicity cores in our sample, it does not give us a sufficient understanding of the nature of the galaxies in the sample.

\subsection{Galaxy pair separation}
Previous studies investigating the MZR or FMR for interacting galaxies have found that close galaxy pairs have a lower metallicity compared to galaxy pairs with a greater projected separation \citep{10.1111/j.1745-3933.2008.00466.x, 2020MNRAS.494.3469B}.
It should be noted that metallicity will have a scatter as at any given distance a galaxy pair can be at a number of different stages of the merger process.
\begin{figure*}[htp]
    \centering
    \begin{tabularx}{\textwidth}{X}
    \includegraphics[width=\textwidth,keepaspectratio]{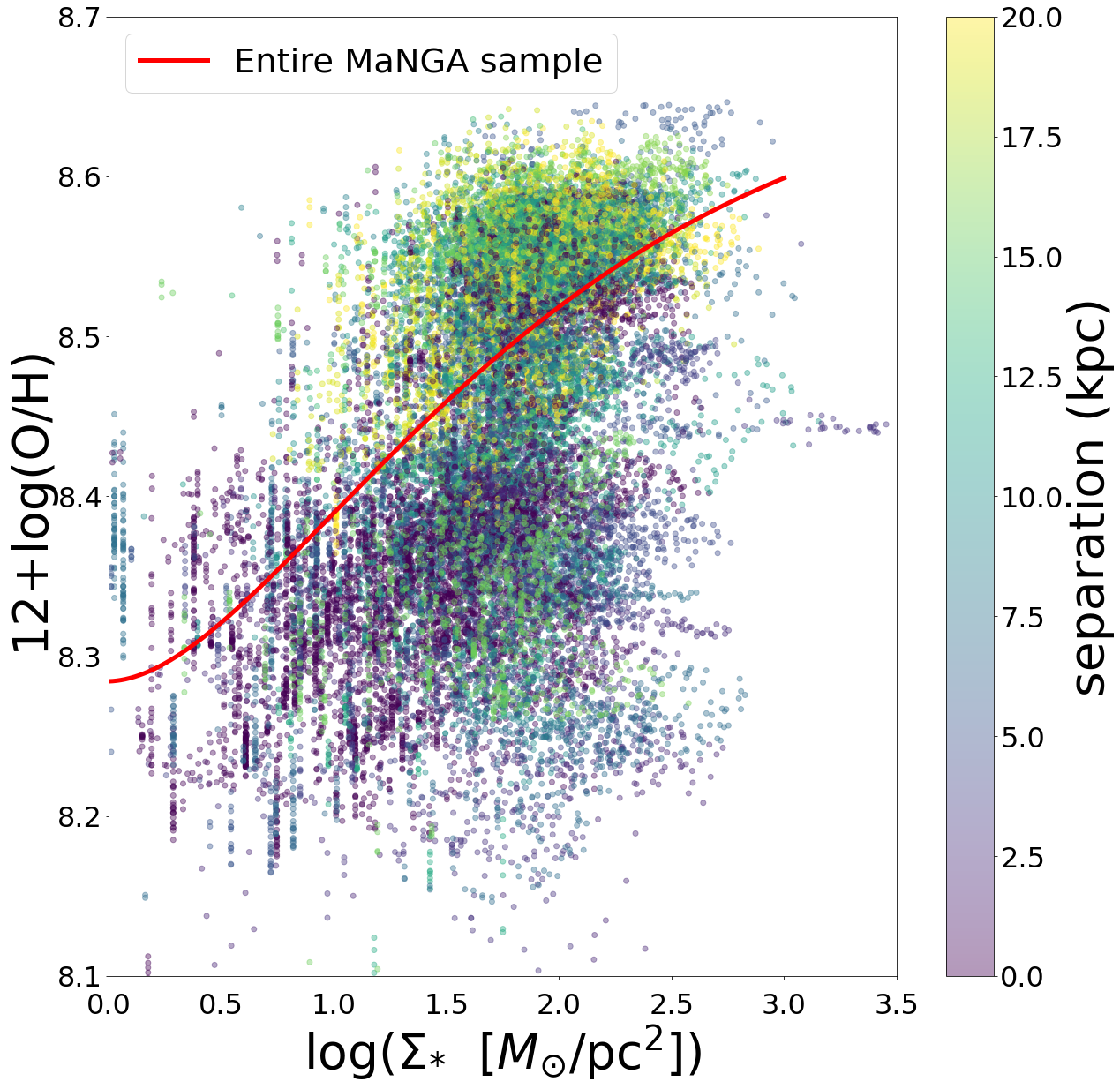}
    \end{tabularx}
    \caption{Distribution of oxygen abundance as a function of stellar mass surface density for all star-forming spaxels in the 298 galaxy pairs in the MaNGA survey, colour-coded by galaxy pair separation. All spaxels over 1 $R_{\textrm{eff}}$ are masked.}
    \label{fig:sep}
\end{figure*}

Figure \ref{fig:sep} plots all spaxels within 1.5 $R_{\textrm{eff}}$ colour-coded by the projected separation of the galaxy pair the spaxel belongs to.
We find that spaxels that have a gas phase metallicity consistent with that of the MZR, with increased stellar mass resulting in increased metallicity, likely belong to a galaxy pair with a greater projected separation.
We also find that spaxels with diluted metallicity belong to galaxies with lower projected separation.
In other words, closer galaxy pairs are more diluted, particularly in their nuclear regions, as can be seen from the previous section.
The closest galaxy pairs experiencing a dilution in metallicity is consistent with previous works that studied the MZR or FMR for galaxy pairs \citep{Rupke_2010, 2020MNRAS.494.3469B, 2021MNRAS.501.2969G}.

Galaxy pairs with a close separation ($<5$ kpc separation) can indicate a merger at a number of different stages.
Galaxies can be near the first pericentre passage or approaching coalescence. 
At both of these stages, galaxies are experiencing metallicity dilution in circumnuclear regions.
In the first case, the primary galaxy experiences an inflow of low-metallicity gas from the secondary galaxy, and the gas phase metallicity abundance is diluted \citep{2004ApJ...613..898T, 2012ApJ...746..108T, Montouri2010}.
In the second case, when galaxies are approaching coalescence, strong gas inflows are observed, resulting in a dilution of nuclear metallicities \citep{2012ApJ...746..108T}.
Close galaxy pairs can also be post-coalescence and experience enrichment from supernova ejecta \citep{Montouri2010}. 

Galaxy pairs with a projected separation of $>20$ kpc can be at the first encounter, before the first pericentre passage.
At this stage, little metallicity evolution is observed \citep{Rupke_2010, 2012ApJ...746..108T}, so the metallicity abundance would be expected to be in accordance with the MZR. The pair can also be in a state after the first pericentre passage, where star formation events enhance the nuclear metallicity \citep{2012ApJ...746..108T}.
However, the pair could also be in a period after the first pericentre passage and in the midst of separation before final coalescence, when inflow events will dilute the circumnuclear metallicity \citep{Montouri2010}.

We note that the majority of our galaxies are within 20 kpc in projected separation; as such, they would all be classified under the general term `close pairs' and commonly placed in a single bin in previous literature.
However, we find that there is a possible bimodality dependent on projected separation even with this upper limit of separation. This bimodality is consistent with the metallicity evolution in a galaxy merger.

\subsection{Possible selection effects}
\begin{figure}[h]
    \begin{tabularx}{\textwidth}{X}
    \includegraphics[width=\columnwidth,keepaspectratio]{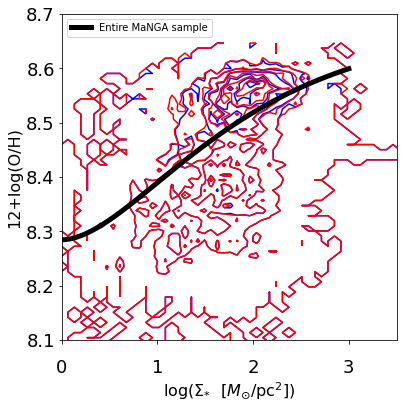}
    \end{tabularx}
    \caption{Contours of the redshift distribution of our sample. The red and blue contours represent spaxels below and above redshift 0.1.}
    \label{fig:redshift}
\end{figure}

Figure \ref{fig:redshift} shows the contours of the redshift distribution of our sample. 
We find that spaxels that belong to galaxies with $z>0.1$ show a metallicity enrichment, whereas spaxels that belong to galaxies with $z<0.1$ are more scattered throughout the MZR.
We note that the lack of metallicity-diluted galaxies at $z>0.1$ is possibly due to selection effects.
Following the luminosity-metallicity relation \citep{1989ApJ...347..875S, 2001A&A...374..412P, 2009A&A...505...63G}, lower-metallicity galaxies have lower intrinsic luminosities, resulting in observational limits of their detection.
Additionally, merger classifications are incomplete at higher redshifts \citep{Huertas_Company_2015}.

%


\section{Conclusions}

\label{section:Conclusion}
In this work we have investigated the gas phase metallicity ($12 + \log (\textrm{O/H})$) as a function of stellar mass, or the MZR, of star-forming spaxels in MaNGA galaxy pairs identified using visual and kinematic features.
Our main findings include the following:
   \begin{enumerate}
      \item We find a bimodality -- two peaks in the distribution of spaxels in the mass-metallicity space for galaxy pairs -- a feature that is not present in the MaNGA sample as a whole. This bimodality was not observed in the study by \citet{2020MNRAS.499.4370M}, a previous spatially resolved study on perturbed galaxies.
      \item The spaxels at the peaks correspond to spaxels in the cores of the galaxy pairs, indicating both metallicity enrichment and dilution in circumnuclear regions.
      \item Galaxy pairs with closer separations showed a tendency to display metallicity dilution, whereas galaxy pairs with greater separations showed a metallicity enrichment. This is likely an indicator of metallicity evolution during the galaxy merger process.
   \end{enumerate}

Previous studies on the global MZR of interacting galaxies have found that there is a metallicity dilution present for interacting galaxies compared to isolated galaxies of similar stellar masses.
Our results show that a metallicity dilution can be observed for interacting galaxies at the local level; however, there is a bimodality that is likely attributable to galaxy separation.

In future works, we plan to investigate the star formation rate of this galaxy sample and investigate the effects of galaxy interaction on the FMR for MaNGA galaxy pairs.

\begin{acknowledgements}
     We would like to thank the anonymous referee for their comments in improving this manuscript.
     This work has been supported by the Japan Society for the Promotion of Science (JSPS) Grants-in-Aid for Scientific Research (19H05076 and 21H01128). 
This work has also been supported in part by the Sumitomo Foundation Fiscal 2018 Grant for Basic Science Research Projects (180923), and the Collaboration Funding of the Institute of Statistical Mathematics ``New Development of the Studies on Galaxy Evolution with a Method of Data Science''. 
     
\end{acknowledgements}

%
%

\bibliography{References}
\bibliographystyle{aa}
\appendix
\section{MZR of all MaNGA star-forming spaxels}
\label{appendix: A}
\noindent
\begin{minipage}{\textwidth}
    \centering
    \begin{tabularx}{\textwidth}{X}
    \includegraphics[width=\textwidth,keepaspectratio]{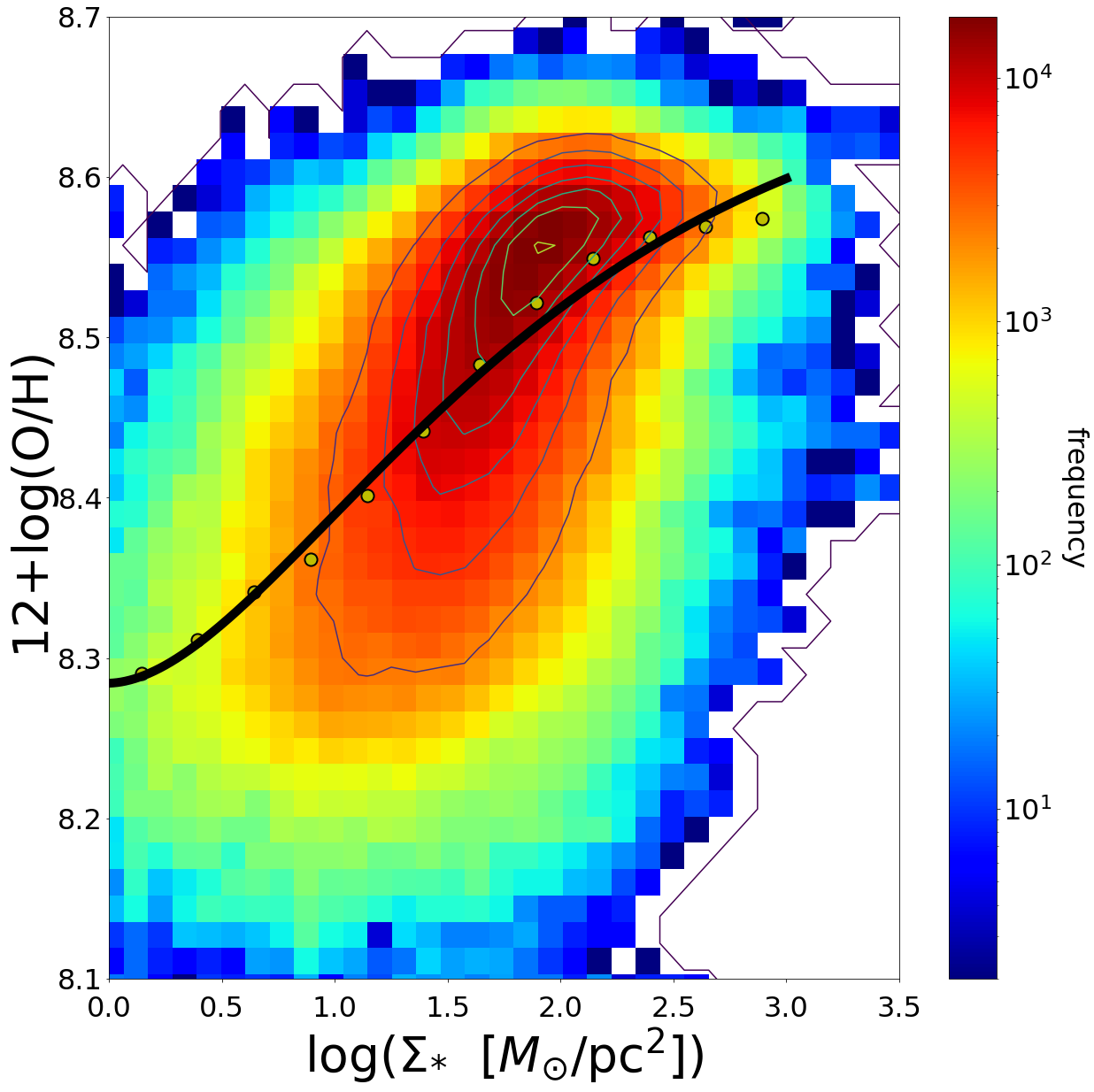}
    \end{tabularx}
    \captionof{figure}{Distribution of oxygen abundance as a function of stellar mass surface density for all star-forming spaxels from all galaxies in the MaNGA survey. The colour bar indicates the number of spaxels per bin in the $\Sigma_{*}-\mathrm{Z}$ space. The black curve is the best-fit line following \citet{2013A&A...554A..58S}.}
    \label{figure: allmangamzr}    
\end{minipage}

To obtain the curve of best fit, we found the median value in differing mass density bins and, following \citet{2013A&A...554A..58S}, used the following fitting function:
\begin{equation}
    y=a+b(x-c)e^{-(x-c)}
,\end{equation}
with $y=12 +  \log \text{(O/H)}$ and $x$ the logarithm of the stellar mass surface density.
From the best-fit line, we find the coefficients for the above function for all star-forming spaxels in the MaNGA survey to be $a=8.68\pm0.01$, $b=-1.07\pm0.04$, and $c=-1.01\pm0.12$.

\end{document}